\documentstyle[aps,preprint,floats,psfig]{revtex}
\begin{document}
\newcommand{\beq}{\begin{equation}}
\newcommand{\eeq}{\end{equation}}
\draft
\title  {Josephson effect test for triplet pairing symmetry}
\author{ N. Stefanakis}
\address{ Department of Physics, University of Crete,
	P.O. Box 2208, GR-71003, Heraklion, Crete, Greece}
\date{\today}
\maketitle

\begin{abstract}

The critical current modulation and the spontaneous flux of the vortex states in corner Josephson junctions between Sr$_2$RuO$_4$ and 
a conventional $s$-wave superconductor
are calculated
as a function of the crystal orientation, and the magnetic field.
For Sr$_2$RuO$_4$ we assume two nodeless $p$-wave pairing states. 
Also we use the nodal $f$-wave states $B_{1g}\times E_u$ and 
$B_{2g} \times E_u$, and one special $p$-wave state having line nodes. It is seen that the critical current depends solely on the topology of the gap.
\end{abstract}

\newpage
\section{Introduction}

During recent years, a renewed interest has been realized 
for unconventional superconductivity in Sr$_2$RuO$_4$ (SRO)
\cite{maeno}. This 
material has a layered structure and is a two dimensional Fermi liquid,
with similarities to $^3$He, that shows $p$-wave superfluidity. 
Knight-shift measurements suggest 
that unconventional superconductivity with pairing in the triplet 
channel is realized \cite{ishida}. 
The time-reversal symmetry 
is violated for this state. This is consistent 
with the discovery of internal magnetic fields in the 
superconducting phase by $\mu$SR measurement \cite{luke}.
Also specific-heat measurements support the scenario of
line nodes within the superconducting gap 
as in the high-$T_c$ cuprate superconductors \cite{nishizaki}.
Moreover thermal-conductivity measurements 
indicate the presence of horizontal lines of nodes \cite{izawa}.
The argument for unconventional superconductivity in SRO is further 
supported by the variation of $T_c$ with nonmagnetic impurities
\cite{mackenzie}. 

The tunneling effect in ferromagnet-triplet-superconductor 
junctions can be used to determine the nodal structure of the 
pair potential
and to distinguish the proposed pairing states. 
When the interface is normal to the $ab$ plane, for two-dimensional 
order parameters,
tunneling peaks are formed due to bound states 
that depend both on the surface orientation and the quasiparticle trajectory 
angle \cite{stefan2}. 
For three dimensional order parameters the tunneling spectra is sensitive 
to the orientation of the interface \cite{stefan3}.

The Josephson effect has been observed experimentally in 
bilayer junctions Pb/Sr$_2$RuO$_4$/Pb, where a sharp drop 
of the critical current $I_c$ is observed below the transition 
temperature of Sr$_2$RuO$_4$ \cite{jin1}. 
This anomalous effect originates from the competition between the 
positive Josephson coupling strength between the two singlet 
superconductors and the negative one due to the 
singlet-/triplet-superconductor coupling \cite{yamashiro}.
Also in Ref. \cite{jin2} it is
argued that the 
Josephson coupling between a conventional $s$-wave superconductor 
In and Sr$_2$RuO$_4$ is allowed in the in-plane direction, 
but not in the $c$ axis. The conclusion is that the pairing state 
in that case has to be a nodeless $p$-wave state.

In order to distinguish the pairing states with a two-dimensional 
order parameter,
in the present paper we study the static properties of a two-dimensional 
corner junction 
between SRO and a 
conventional $s$-wave superconductor. 
This is done by mapping the two dimensional (2D)
junction into two single junctions, connected in parallel,
by introducing an extra relative phase in one part of this junction.
The spontaneous flux and the critical current modulation 
of the vortex states with the junction 
orientation are calculated by solving numerically 
the sine-Gordon equation. 
For SRO we shall assume
three possible pairing states of two-dimensional
order parameters,
having line nodes within the RuO$_2$ plane, that
break the time-reversal symmetry.
The first two are the 2D, $f$-wave states
proposed by Hasegawa {\it et al.},
\cite{hasegawa} having $B_{1g}\times E_u$ and $B_{2g}\times E_u$ symmetry,
respectively.
The other one is called the nodal $p$-wave state and has been
proposed by Dahm {\it et al.} \cite{dahm}, where
the pairing potential has the form
${\bf d}=\Delta_0\hat{\bf z}[\sin(k_xa)+i\sin(k_ya)]$. 
This pairing symmetry has nodes as in the $B_{2g}\times E_u$ case.
Also we will consider two nodeless pairing states. One is
the isotropic $p$-wave state and the other is the
nodeless $p$-wave state initially proposed by
Miyake and Narikiyo \cite{miyake}, both breaking
the time-reversal symmetry. In addition we examine 
briefly several three-dimensional 
order parameters with horizontal lines of nodes.

For each pairing state we derive simple arguments that could be 
used to characterize it.
For example the junction orientation where the $a$,$b$ axes point towards the 
junction interface will give finite critical currents only 
for the isotropic $p$-wave, $B_{1g}\times E_u$ states and zeros 
for the other states. 
Furthermore the suppression of 
the critical currents of the nodal state
$B_{1g}\times E_u$  near the nodes
can be used to distinguish it from the
nodeless isotropic $p$-wave state.
Also the modulation of the Josephson critical current 
with the polar angle $\beta$, 
within the $ac$ plane, will probe the horizontal nodes for 
hypothetical three-dimensional order-parameter symmetries.
The rest of the paper is organized as follows. In Sec. II we
discuss the Josephson effect between a superconductor breaking 
time-reversal symmetry and a conventional 
$s$-wave superconductor. In Sec. III the geometry of the corner junction is 
discussed. In Sec. IV 
we present the magnetic flux for the spontaneous vortex states.
In Sec. V the modulation of the magnetic flux and the critical 
currents with the orientation
are presented. 
In Sec. VI several three-dimensional order parameters are considered.
In Sec. VII the modulation of the vortex states 
with the magnetic field is discussed. 
In Sec. VIII a connection with the experiment is made.
Finally, a summary and discussion are presented in the last section.

\section{Josephson effect between SRO and a conventional $s$-wave 
superconductor} 

We consider the junction shown in Fig. \ref{fig1.fig}(a), between a
superconductor $A$ with a two-component 
order parameter $n_A$, and a
superconductor $B$ with order parameter $n_B$. 
The two superconductors 
are separated by an intermediate layer.
The bulk order parameters 
$n_A$ and $n_B$, 
near the interface, can be written as
\beq 
  n_A = \left\{ 
    \begin{array}{ll}
      \widetilde{n}_{A_x}e^{i \phi_{A_x}}, & \\
      \widetilde{n}_{A_y}e^{i \phi_{A_y}} & 
    \end{array}~~~\label{pnip}
\right.
\eeq

and 

\beq 
  n_B =
      \widetilde{n}_Be^{{\it i} \phi_B}
   ,~~~\label{pnis}
\eeq
where $\phi_{A_x}, \phi_{A_y} $ are the phases of the
components of the order parameter $n_A$,
and $\phi_B$ is the phase of the
superconductor $B$. $\widetilde{n}_{A_x}$, $\widetilde{n}_{A_y}$ are 
the magnitudes of the components of the order parameter $n_A$.
Then the supercurrent density can be written as 
\beq
J=J_{BA_x}\sin(\phi_B-\phi_{A_x})+ 
J_{BA_y}\sin(\phi_B-\phi_{A_y}),~~~\label{pj}
\eeq
where 
\beq 
\begin{array}{ll}
J_{BA_x} \sim &  \widetilde{n}_{A_x} \widetilde{n}_B,\\
J_{BA_y} \sim &  \widetilde{n}_{A_y} \widetilde{n}_B.
\end{array}~~~\label{pjkl}
\eeq

We have to comment here that the coupling between a triplet and 
a singlet superconductor is forbidden by the orthogonality 
of the spin part of their wave functions. However, in our situation
there exists a finite overlap due to spin-orbit coupling
\cite{honerkamp}. In the presence of spin-orbit coupling 
the Cooper pairs of different parities will be mixed at the 
interface between the singlet and the triplet superconductor 
resulting in a direct Josephson coupling between them \cite{larkin}. 
We define $\phi=\phi_B-\phi_{A_x}$ as the relative phase difference
between two superconductors. 
We consider the case where the intrinsic phase difference 
within superconductor $A$ is 
$\phi_{A_y}-\phi_{A_x}=\pi/2$. In this case the order parameter 
is complex and breaks 
the time-reversal symmetry.
Then the supercurrent density can be written as: \cite{zhu}

\beq
J(\phi)=\widetilde J_c \sin(\phi+ \phi_c),~~~\label{pjphidis}
\eeq
with 

\beq 
  \phi_c = \left\{ 
    \begin{array}{ll}
      \tan^{-1}\frac{-J_{BA_y}}{J_{BA_x}} & J_{BA_x} > 0, \\ 
      \pi + \tan^{-1}\frac{-J_{BA_y}}{J_{BA_x}} & J_{BA_x} < 0,  
    \end{array}~~~\label{phic}
\right.
\eeq
and
\beq
\widetilde J_c=\sqrt{J_{BA_y}^2+J_{BA_x}^2}.~~~\label{jcp}
\eeq

We consider the following cases for the pairing state of 
SRO having line nodes.

(a) In the first 2D $f$-wave state $B_{1g}\times E_u$,
the magnitudes of the order parameters $\widetilde{n}_{A_x}$ 
and $\widetilde{n}_{A_y}$ in Eq. (\ref{pjkl}) are defined as
$\widetilde{n}_{A_x}=\cos(2\alpha)\cos(\alpha)$, 
$\widetilde{n}_{A_y}=\cos(2\alpha)\sin(\alpha)$.
In these formulas
$\alpha$ denotes the angle between the interface
and the $a$ axis of the crystal.
This state has nodes at the same points as in the $d_{x^2-y^2}$-wave
case.

(b) For the second 2D $f$-wave state $B_{2g}\times E_u$, 
$\widetilde{n}_{A_x}=\sin(2\alpha)\cos(\alpha)$, 
$\widetilde{n}_{A_y}=\sin(2\alpha)\sin(\alpha)$.
This state has nodes at $0, \pi/2, \pi$, and $3\pi/2$, and has also
been studied by Graf and Balatsky \cite{graf}.

(c) This is the case of a nodal $p$-wave superconductor,
$\widetilde{n}_{A_x}=1/s_M \sin[\pi \cos(\alpha)]$, 
$\widetilde{n}_{A_y}=1/s_M \sin[\pi \sin(\alpha)]$.
We use here the same normalization proposed by Dahm {\it et al.}, \cite{dahm}
$s_M=\sqrt{2}\sin(\pi/\sqrt{2})=1.125$, where the
Fermi wave vector is chosen as $k_Fa=\pi$, in order to have
a node of the order parameter in the Fermi surface.
This state has nodes as in the $B_{2g}\times E_u$ state.
The corresponding nodeless form was initially
proposed by Miyake, and Narikiyo
\cite{miyake} and will be considered as a separate case.
Also we will examine the following two pairing symmetries,
which are nodeless

d) In the case of a nodeless $p$-wave superconductor, 
proposed by Miyake and Narikiyo
\cite{miyake}
$\widetilde{n}_{A_x}=1/s_M \sin[R \pi \cos(\alpha)]$, 
$\widetilde{n}_{A_y}=1/s_M \sin[R \pi \sin(\alpha)]$, 
$s_M=\sqrt{2}\sin(\pi/\sqrt{2})=1.125$, and $R=0.9$.
This state does not have nodes.

e) In the isotropic $p$-wave case, 
$\widetilde{n}_{A_x}=\cos(\alpha)$, 
$\widetilde{n}_{A_y}=\sin(\alpha)$.
This pairing state does not have nodes either.

\section{The junction geometry} 
\begin{figure}
  \centerline{\psfig{figure=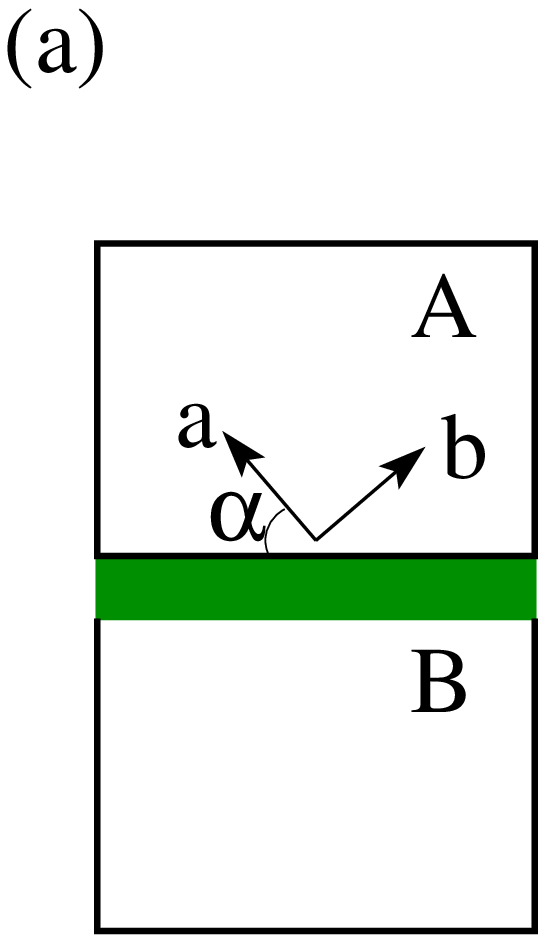,width=3.cm,angle=0}}
  \centerline{\psfig{figure=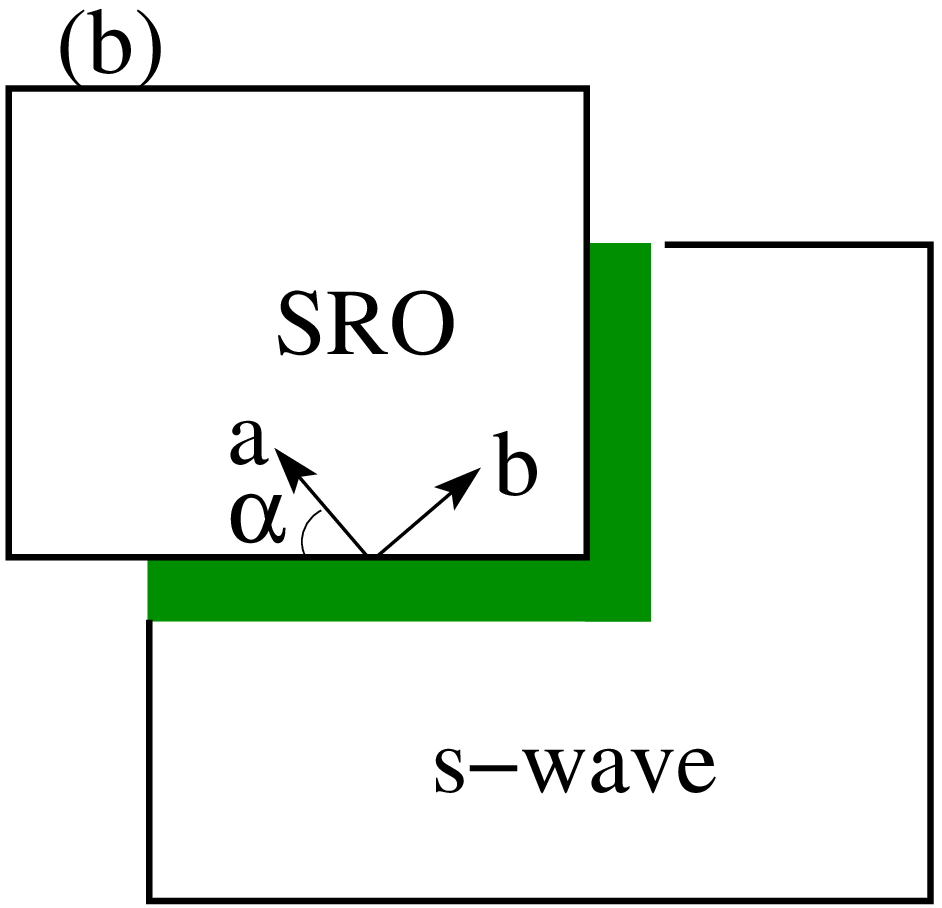,width=5.5cm,angle=0}}
  \centerline{\psfig{figure=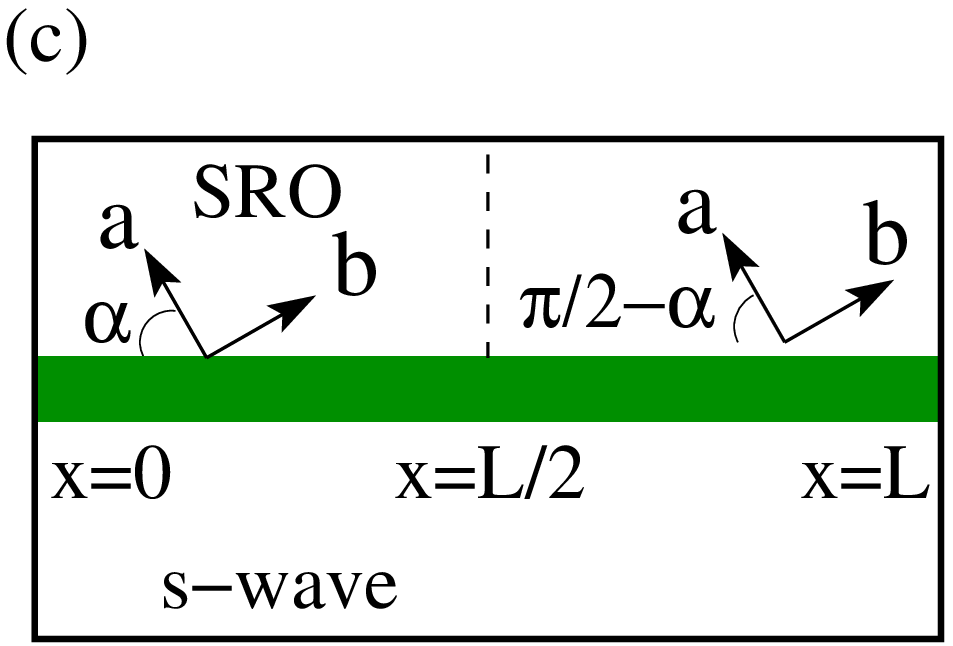,width=5.5cm,angle=0}}
  \caption{(a) A single Josephson junction between a triplet 
superconductor $A$ with 
broken time-reversal pairing symmetry and 
$B$, with $s$-wave symmetry. The angle between the 
crystalline $a$ axis of $A$ and the junction interface is 
$\alpha$. (b) The geometry of the corner junction between 
superconductors SRO and $s$ wave.
(c) The mapping of the corner junction geometry into the 
one-dimensional axis of length $L$.
For $0<x<L/2$ the angle between the $a$ axis and the interface is 
$\alpha$ while for $L/2<x<L$ the corresponding angle is $\pi/2-\alpha$.}
\label{fig1.fig}
\end{figure}

We consider the corner junction shown in Fig. \ref{fig1.fig}(b) between 
a triplet superconductor breaking the time-reversal symmetry 
and an $s$-wave superconductor.
In the following the Josephson coupling is restricted in the 
$ab$ plane since we assume 2D order parameters.
If the angle of the $a$ axis with the interface in the $x$ direction 
is $\alpha$, then the corresponding angle in the $y$ direction 
will be $\pi/2-\alpha$. Using this constraint we map the two segments 
of this corner junction, each of length $L/2$,
into a one-dimensional axis shown 
in Fig. \ref{fig1.fig}(c).
In this case the two-dimensional junction can be considered as being 
made of two one-dimensional junctions described in Sec. II 
connected in parallel. 
The characteristic phases of the two parts of the one-dimensional junction 
$\phi_{c1}$ and $\phi_{c2}$ depend upon the angle $\alpha$.
We call this junction frustrated since the two segments have different 
characteristic phases $\phi_{c1}$ and $\phi_{c2}$. 
The fabrication details of 
corner junctions or SQUID's, between sample faces at different angles,
can be found in Refs. \cite{vanh} and \cite{yanoff}.

The superconducting phase difference $\phi$ across the junction is 
then the solution of the sine-Gordon equation
\beq 
  \frac{d^2 { \phi}(x)}{dx^2} = \widetilde J_c\sin[{
\phi(x)+\phi_c(x)}] - I^{ov},~~~\label{eq01} 
\eeq 
with the boundary conditions
\begin{equation}
  \frac{d \phi}{dx}|_{x=0,L} = H
,~~~\label{eqbc}
\end{equation}
where $\phi_c(x)=\phi_{c1} (\phi_{c2})$ in the left (right) part of the
junction, and $I^{ov}$ is the bias current in the overlap geometry. 
The length $x$ is scaled in units of
the Josephson penetration depth given by
\[
\lambda_J=\sqrt{\frac{\hbar c^2}{8\pi e d J_{c0}}},
\]
where $d$ is the sum of the $s$-wave, and mixed wave, in-plane
penetration depths plus the thickness of the
insulator layer.
The different solutions obtained from
Eq. (\ref{eq01}) are classified with their magnetic-flux content
\begin{equation}
  \Phi = \frac{1}{2 \pi}  (\phi_R-\phi_L) ,~~~\label{phi}
\end{equation}
where $\phi_{R(L)}$ is the value of $\phi$ at the right (left) edge of
the junction, in units of the flux quantum
$\Phi_0= hc/2e$.

To check the stability of solutions we consider small perturbations 
$u(x,t)=u(x)e^{st}$ on the static solution $\phi(x)$, and linearize 
the time-dependent sine-Gordon (s-G) equation to obtain  
\beq
  \frac{d^2 v(x)}{dx^2} + \widetilde J_c\cos[{
\phi(x)+\phi_c(x)}]v(x) = \lambda v(x), ~~~\label{eqstab}
\eeq
where $\lambda=-s^2$, with the boundary conditions
\begin{equation}
  \frac{dv}{dx}|_{x=0,L} = 0
.~~~\label{eqbclambda}
\end{equation}
It is seen that if the eigenvalue equation has a 
negative eigenvalue the static solution $\phi(x)$ is unstable.

\section{Spontaneous vortex states}
First let us consider the case where the two
one-dimensional junctions between a triplet superconductor breaking
the time-reversal symmetry and an s-wave superconductor, each of length $L/2$,
described in Sec. II, are uncoupled.
Then for $0<x<L/2$ the stable solutions for the s-G equation are
$\phi(x)=-\phi_{c1}+2n_1 \pi$, where $n_1=0,\pm 1,\pm 2,...$,
while for
$L/2<x<L$ the stable solutions for the s-G equation are
$\phi(x)=-\phi_{c2}+2n_2 \pi$, where $n_2=0,\pm1,\pm2,...$, and
$\phi_{c1}$, $\phi_{c2}$ are the relative phases in
each part of the junction due to different orientations.
When the frustrated junction is formed, and we consider the
above junctions in parallel, the phase $\phi$ is forced to
change around $x=L/2$ to connect these stable solutions.
This variation of the phase $\phi$ along the junction
describes the Josephson vortices.
The flux content of these states (in units of $\Phi_0$) is \cite{sigrist}
\begin{equation}
\Phi=[\phi(L)-\phi(0)]/2\pi =
(-\phi_{c2}+\phi_{c1}+2n\pi)/2\pi,~~~\label{con}
\end{equation}
where the $n$ value ($n=n_1-n_2=0,\pm 1,\pm 2,...$)
distinguishes between solutions with
different flux content. We will
concentrate on solutions called modes with minimum flux content, i.e.,
$n=0, 1, -1$.
Their magnetic flux in terms of $\phi_{c1}, \phi_{c2}$
is shown in Table \ref{n=01_1}.
Generally the flux content is fractional, i.e., is neither integer
nor half-integer, as a consequence of the broken time-reversal symmetry
of the problem.

\section{Spontaneous magnetic flux and critical currents} 
In this section we will consider the magnetic flux and critical current
modulation with the azimuthal angle $\alpha$, for the 
various two-dimensional pairing states for the SRO.
In general the flux quantization and the critical currents 
are influenced by the nodal structure of the order 
parameter. The critical current $I_c^{ov}$ which depends mainly on 
the magnitude of the order parameter vanishes at the nodes 
of the order parameter. 
The magnetic flux is determined from $\phi_{c1},\phi_{c2}$, 
and has almost the same structure in the isotropic $p$-wave,
$B_{1g}\times E_u$ and 
$B_{2g}\times E_u$ pairing states, while it is different 
in the nodal $p$-wave and the nodeless $p$-wave cases.
   
\subsection{Isotropic $p$ wave}
\begin{figure}
 \centerline{\psfig{figure=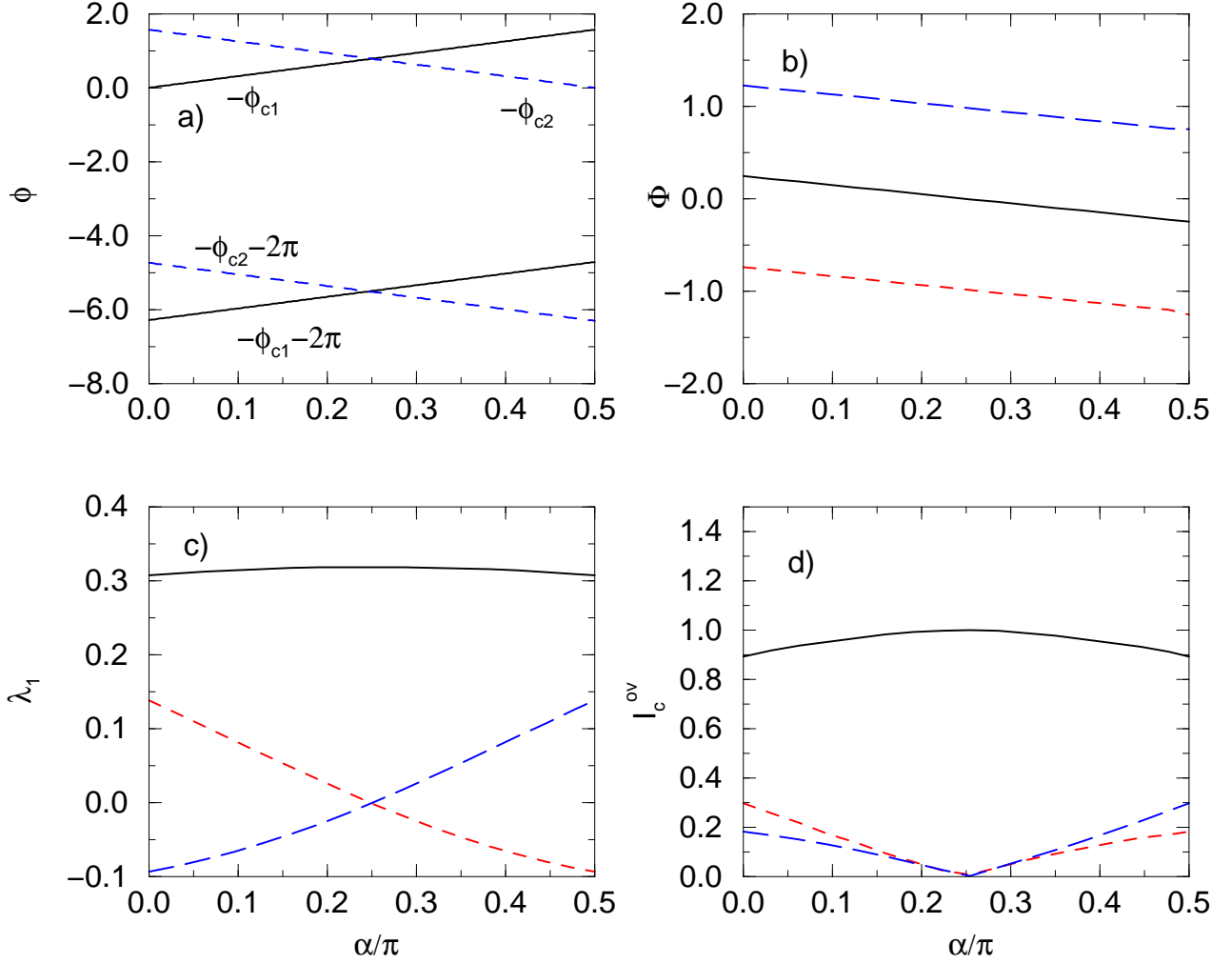,width=17cm,angle=0}}
\caption{
(a) The variation of $\phi_{c1}$, $\phi_{c2}$
with the angle
$\alpha$. For each value of $\alpha$ there exist three vortex 
solutions. 
(b) The spontaneous magnetic flux $\Phi$ as a function of the 
angle $\alpha$, for the $0,-1,1$ mode solutions as solid, 
dashed, and long-dashed lines respectively.
(c) The lowest eigenvalue $\lambda_1$ as a function of the 
angle $\alpha$, for the $0,-1,1$ mode solutions. 
(d) Overlap critical current density $I_c^{ov}$ versus 
the angle $\alpha$ 
for a junction of length 
$L=10$. 
The pairing symmetry of the triplet 
superconductor is isotropic $p$ wave.
}
\label{iso.fig}
\end{figure}

For the isotropic $p$-wave case we present in Fig. 
\ref{iso.fig}(a) the variation of $\phi_{c1}, \phi_{c2}$, with 
the interface orientation angle $\alpha$. 
The magnetic flux can be calculated analytically for the 
spontaneous vortex states $0,-1,1$. The corresponding 
expressions are seen in Table \ref{isotable}.
The spontaneous magnetic flux as a function of the angle $\alpha$ 
is presented in Fig. \ref{iso.fig}(b). At $\alpha=0$ we have the 
existence of three possible spontaneous vortex states. 
As we change the angle $\alpha$, the magnetic flux changes, 
keeping the distance between the modes at a constant value 
equal to a single flux quantum. 
The modes for $\alpha=0$ are displaced to quadratic values 
[i.e., $(1/4+n)\Phi_0$] of the 
magnetic flux in contrast to the $s$-wave case where the flux 
is quantized in integer values $0,-1,1$ \cite{owen}. 

\begin{figure}
 \centerline{\psfig{figure=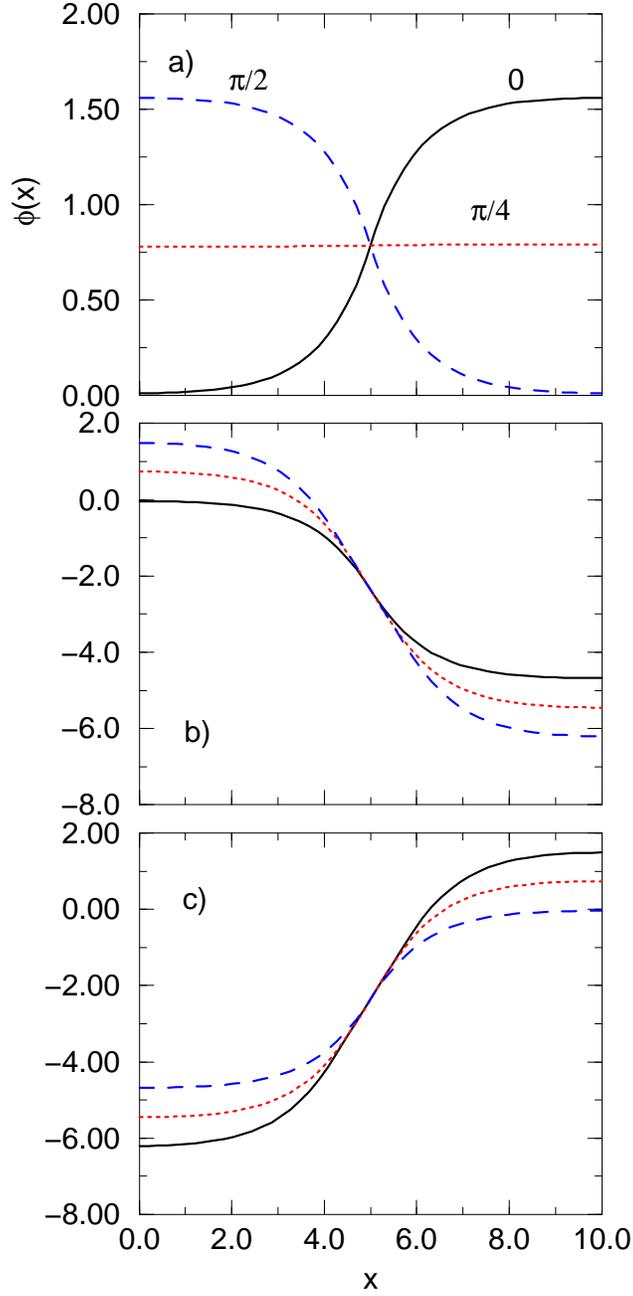,width=8.5cm,angle=0}}
\caption{
The phase distribution of the vortex solutions (a) $n=0$, 
(b) $n=-1$, and (c) $n=1$, 
at $\alpha= 0$, $\pi /4$, $\pi /2$;
for a corner junction of isotropic $p$-wave and $s$-wave
superconductors, with length
$L=10$, and zero overlap external current $I^{ov}=0$.
}
\label{phase.fig}
\end{figure}

In particular the vortex solution in the $n=0$ mode (solid line) contains
$1/4$ of a fluxon for $\alpha=0$, and as we increase the
angle $\alpha$ towards $\pi /4$ it continuously reduces its flux,
i.e., it becomes flat exactly at $\alpha=\pi /4$ and then it
reverses its sign and becomes an antivortex with exactly opposite
flux content at $\alpha=\pi /2$ from that at $\alpha=0$.
In addition we have plotted in Fig. \ref{phase.fig}(a)
the phase distributions for the mode $n=0$ in different
orientations $\alpha=0, \pi /4, \pi /2$. The transition from the vortex
to the antivortex mode as the orientation changes is clearly
seen in this figure.
Note that the solutions in this mode remain stable for all
junction orientations. This is seen in Fig. \ref{iso.fig}(c) where we plot
the lowest eigenvalue ($\lambda_1$) of the linearized eigenvalue
problem as a function of the angle $\alpha$.
We see that $\lambda_1>0$, denoting stability for all values
of the angle $\alpha$ in this mode.

Let as now examine the solution in the $n=-1$ mode [dashed line
in Fig. \ref{iso.fig}(b)]. We see that at
$\alpha=0$ it has negative flux, equal to 
$-0.75$, and is stable. As we increase the angle $\alpha$
it decreases its flux to a full antifluxon when the
orientation is $\pi /4$ and then to flux of
$-1.25$ when $\alpha$ reaches $\pi /2$.
As seen in Fig. \ref{iso.fig}(c) this solution becomes unstable at 
$\alpha= \pi /4$ where a full antifluxon enters the junction.
The phase distribution at this point is 
seen in Fig. \ref{phase.fig}(b) (dotted line).

Finally the solution in the $n=1$ mode [long-dashed line in Fig. 
\ref{iso.fig}(b)] contains 
$1.25$ in flux at $\alpha=0$ and is clearly unstable. It becomes stable
at $\alpha=\pi /4$ [see 
Fig. \ref{iso.fig}(c) (long-dashed line)],
where
a full fluxon enters the junction, 
as seen in Fig. \ref{phase.fig}(c) (dotted line).
At $\alpha=\pi /2$ it contains $0.75$ in flux.

\begin{figure}
 \centerline{\psfig{figure=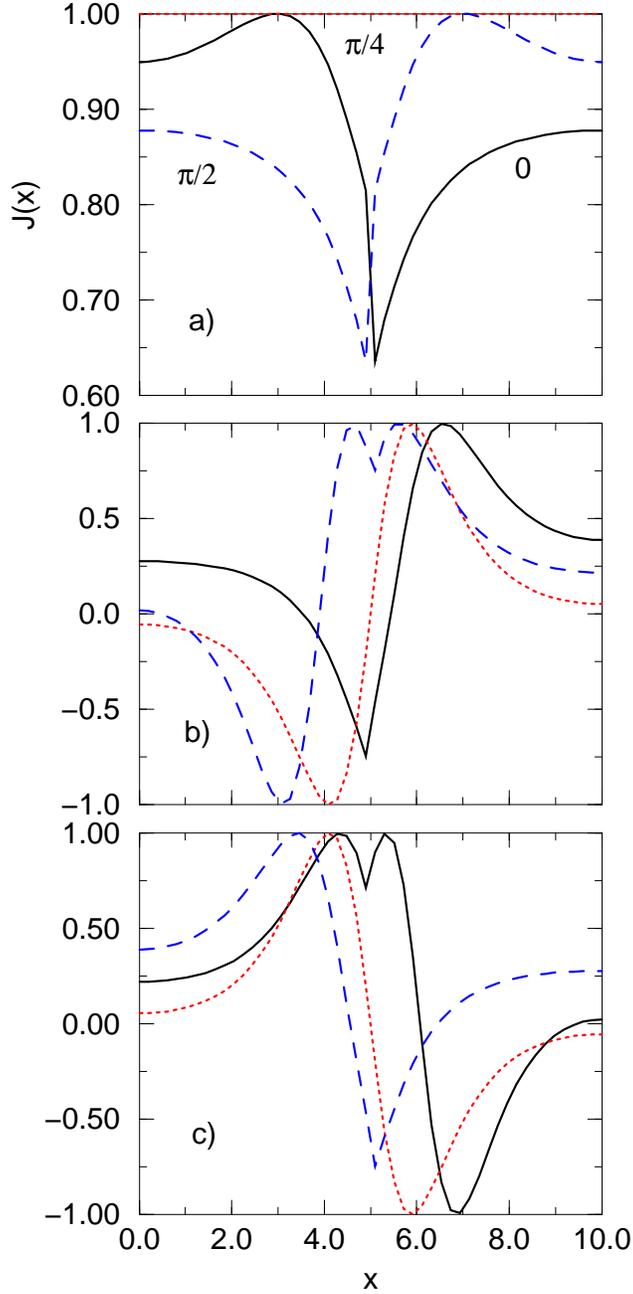,width=8.5cm,angle=0}}
\caption{
The current density distribution $J(x)$ of the vortex solutions (a) $n=0$,
(b) $n=-1$, and (c) $n=1$,
at $\alpha= 0$, $\pi /4$, $\pi /2$;
for a corner junction of isotropic $p$-wave and $s$-wave
superconductors, with length
$L=10$, and maximum overlap external current $I^{ov}$.
}
\label{cdensity.fig}
\end{figure}

In Fig. \ref{iso.fig}(d) we plot the overlap critical current  
per unit length $I_c^{ov}$ as a function
of the angle $\alpha$, at $H=0$, for the $n=0,-1,1$-mode solutions.
The Josephson critical
current density is kept constant with the orientation 
$\widetilde J_c=1$.
Let us consider the situation where the
junction contains a solution in the mode $n=0$, at $\alpha=0$,
when the net current is maximum.
The spatial variation of $\phi$ is described by a fractional vortex
displaced from the
corresponding distribution at zero current which is around $\pi /4$
[see Fig. \ref{phase.fig}(a)].
The current-density distribution as seen in 
Fig. \ref{cdensity.fig}(a) (solid line) corresponds to a positive fluxon,
and at the maximum current is flat, close to unity,
and has a small variation around the junction center giving
rise to the large value of the net current, seen in Fig. \ref{iso.fig}(d).
At $\alpha=\pi/4$ the flat phase distribution corresponding 
to the $n=0$ solution seen in Fig. \ref{phase.fig}(a)
at zero current is displaced towards
the value $3\pi/4$ at the maximum current. 
The current-density distribution as seen in 
Fig. \ref{cdensity.fig}(a) (dotted line) is flat,
giving a net current equal  
to $1$ which is the maximum value over the range $0<\alpha<\pi/2$. 

For the $-1$ solution at $\alpha=0$ 
the phase represents an antifluxon which is 
being displaced at the maximum current 
from the corresponding distribution at zero current seen in 
Fig. \ref{phase.fig}(b). 
The corresponding current density represents an antifluxon,
and has a characteristic variation along  
the junction, 
taking positive and negative values,
as seen in Fig. \ref{cdensity.fig}(b) (solid line), so that 
the net current is suppressed compared to the $0$ mode.
At $\alpha=\pi/4$ just one antifluxon enters the junction 
and the current distribution is symmetric around 
$x=L/2$, as seen in Fig. \ref{cdensity.fig}(b) (dotted line),
leading to zero total current at this angle.

The critical current $I_c^{ov}$ for the $0$ mode 
is finite for each orientation $\alpha$ 
due to the nodeless form of the     
order parameter. This is in marked contrast to what we see in 
the singlet pairing 
states that describe the high-$T_c$ cuprates where the 
critical current $I_c^{ov}$ for the different modes is 
suppressed near $\alpha=\pi/4$ following a similar variation to the 
Josephson critical current density $\widetilde J_c(\alpha)$ \cite{tanaka}.
Also the fractional values of the magnetic flux are originated from the 
two-component order parameter that breaks the time-reversal symmetry.

\subsection{$B_{1g}\times E_u$}

\begin{figure}
 \centerline{\psfig{figure=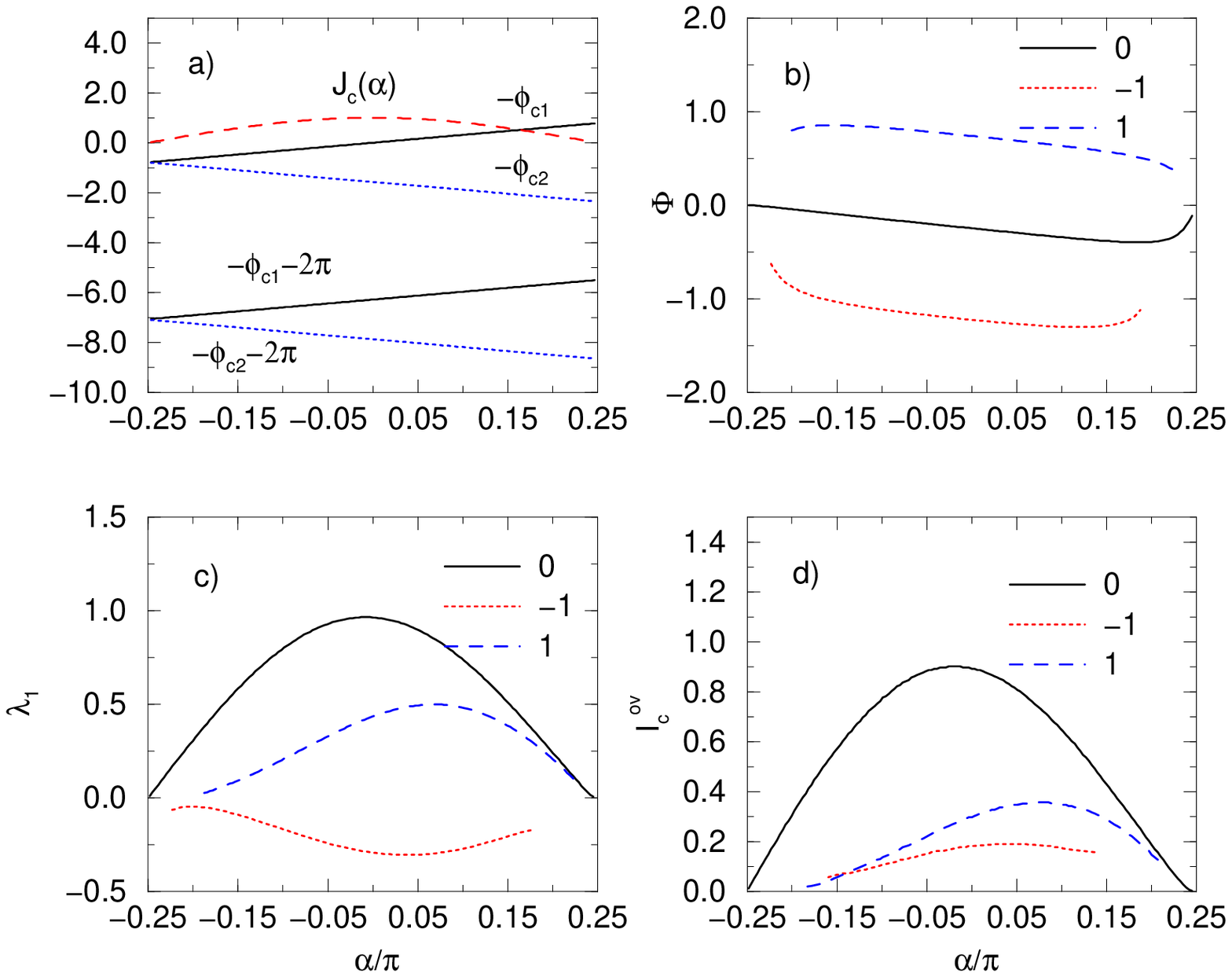,width=17cm,angle=0}}
\caption{
The same as in Fig. 2.
The pairing symmetry of the triplet 
superconductor is $B_{1g}\times E_u$.
}
\label{cos.fig}
\end{figure}

In the $B_{1g}\times E_u$-wave pairing state the 
$\phi_{c1}, \phi_{c2}$ and also the magnetic flux seen in 
Fig. \ref{cos.fig}(a) and Fig. \ref{cos.fig}(b), respectively, have the 
same form as in the isotropic $p$-wave case. 
This happens because the relative phase 
$\phi_c$ in Eq. (\ref{phic}) depends 
on the ratio of the real and imaginary parts of the 
order-parameter magnitude in superconductor SRO.
The $B_{1g}$ component $\cos2\alpha$ multiplies 
both parts and is canceled out. 
On the contrary only the $B_{1g}$ 
part of the order parameter  
contributes to the Josephson critical current density $\widetilde J_c$,
which varies as 
$|\cos(2\alpha)|$ as 
seen in Fig. \ref{cos.fig}(a) (dashed line), while the $E_u$ part 
is canceled. 
It has nodes at $\alpha=\pm\pi/4$.
For this reason we choose the interval $-\pi/4 < \alpha < \pi/4$ 
to study the $B_{1g}\times E_u$ pairing state. 

\begin{figure}
 \centerline{\psfig{figure=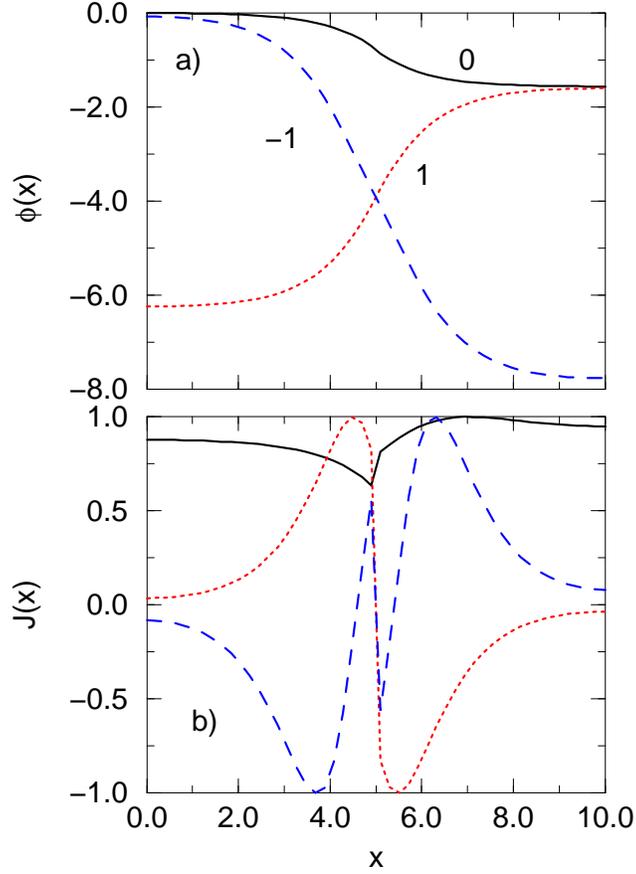,width=8.5cm,angle=0}}
\caption{
(a) The phase distribution of the vortex solutions
at $\alpha=0$, for a junction with length $L=10$, 
and zero external current $I^{ov}=0$.
(b) The corresponding current density at the maximum external 
current.
The pairing symmetry of the triplet 
superconductor is $B_{1g}\times E_u$.
}
\label{cosphase.fig}
\end{figure}

For $\alpha=0$ the mode $n=0$ contains negative magnetic flux 
of $-0.25$ as seen in Fig. \ref{cos.fig}(b), the mode 
$1$ contains positive magnetic flux equal to $0.75$ and is stable, 
while the mode $-1$ has $-1.25$ in flux and is unstable.
The variation of the phase of the vortex solutions at zero current 
and the corresponding current density at maximum current 
for $\alpha=0$ are seen in Fig. 
\ref{cosphase.fig}. Differently from the isotropic $p$-wave case, the mode 
$0$ with negative magnetic flux corresponds to the highest 
critical current.

Close to $\pm \pi /4$ the variation of 
the magnetic flux for the modes $0,-1,1$ 
is influenced from the suppression of the 
$\widetilde J_c$ which goes to zero at $\pm \pi /4$ and the 
magnetic flux acquires some curvature at these points as 
seen in Fig. \ref{cos.fig}(b). 
Due to this abrupt variation the numerical procedure 
to follow modes $-1,1$ stops at the point where the flux 
changes more rapidly.
This is the reason why the magnetic flux for the 
modes $-1,1$ does not reach the nodes of the $\widetilde J_c$
at $\alpha=\pm \pi /4$.

Also the critical currents are influenced by the $\widetilde J_c$. 
The critical current for the mode $0$ vanishes at $\alpha=\pm \pi /4$ 
where the order parameter has nodes.
Also the maximum of the critical current 
tends to move towards 
$\alpha=0$ where $\widetilde J_c$ has its maximum value. 
On the other hand the phase prefers the orientation $\alpha=-\pi /4$ 
where it is flat. 
These two competitive factors shift the maximum current for $n=0$ to the 
left of $\alpha=0$. 
The critical currents for the other two modes are suppressed close
to $\alpha=\pm \pi /4$ where the order parameter has nodes.
In particular for mode $1$ the magnetic flux is smaller compared to 
mode $-1$ and as a consequence the critical current is higher.
In the interval we study, mode $-1$ is unstable while modes 
$1,0$ are stable as seen in Fig.  \ref{cos.fig}(c), where the 
lowest eigenvalue $\lambda_1$ versus $\alpha$ is plotted. 

\subsection{$B_{2g}\times E_u$} 

\begin{figure}
 \centerline{\psfig{figure=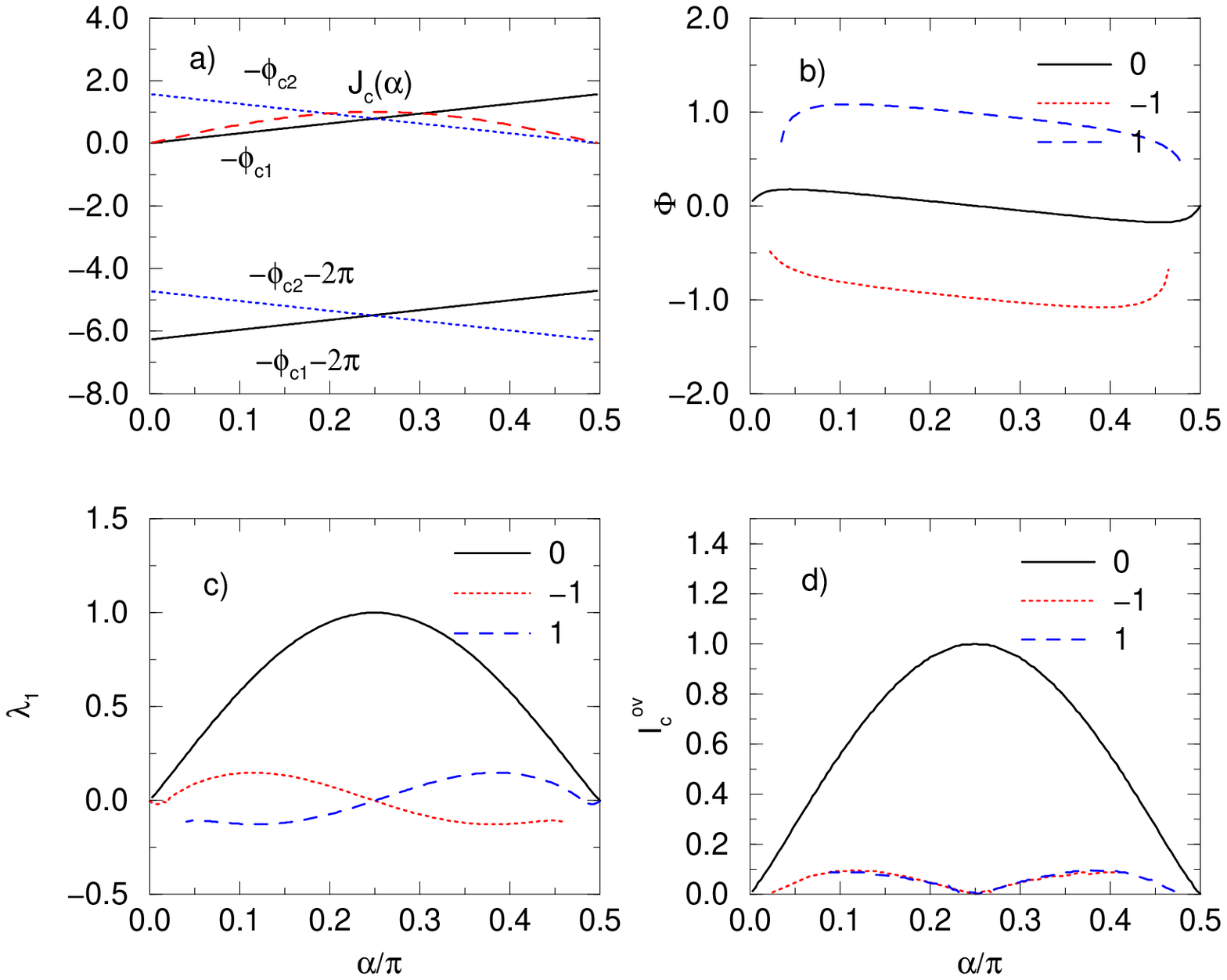,width=17cm,angle=0}}
\caption{
The same as in Fig. 2.
The pairing symmetry of the triplet 
superconductor is $B_{2g}\times E_u$.
}
\label{sin.fig}
\end{figure}

In the $B_{2g}\times E_u$ case the $\phi_{c1}, \phi_{c2}$ 
seen in
Fig. \ref{sin.fig}(a)
have the same form as in the isotropic $p$ wave case. 
Here the $B_{2g}$ factor of the order parameter
$\sin2\alpha$ is canceled out in the calculation of the 
relative phase $\phi_c(x)$. The $B_{2g}$ part of the 
order parameter influences the Josephson critical current 
$\widetilde J_c$, which varies as $|\sin(2\alpha)|$ as 
seen in Fig. \ref{sin.fig}(a) (dashed line) and has nodes at $0, \pi /2$.
Due to this form of the $\widetilde J_c$ we have 
nodes at the critical current $I_c^{ov}$ versus $\alpha$ at the 
points $0, \pi /2$, as seen in Fig. \ref{sin.fig}(d).
The modes $1,-1$ have  zero critical current at the 
angle $\alpha=\pi/4$ where a full fluxon or antifluxon 
enters into the junction. 
In addition they become zero close to $0$ or $\pi /2$ 
due to the vanishing of $\widetilde J_c$ at these angles.
Note that close to $0$ or $\pi /2$ the form of $\widetilde J_c$
influences also the magnetic flux at zero current 
which acquires a small curvature at these points.
The mode 0 
has its maximum close to $\alpha=0$ in the $B_{1g}\times E_u$ state 
while in $B_{2g}\times E_u$
the critical current at $\alpha=0$ becomes zero. 
In the isotropic $p$-wave case the critical current is finite
at $\alpha=0$ 
since the order parameter for this state is nodeless.

\subsection{Nodal $p$ wave}
\begin{figure}
 \centerline{\psfig{figure=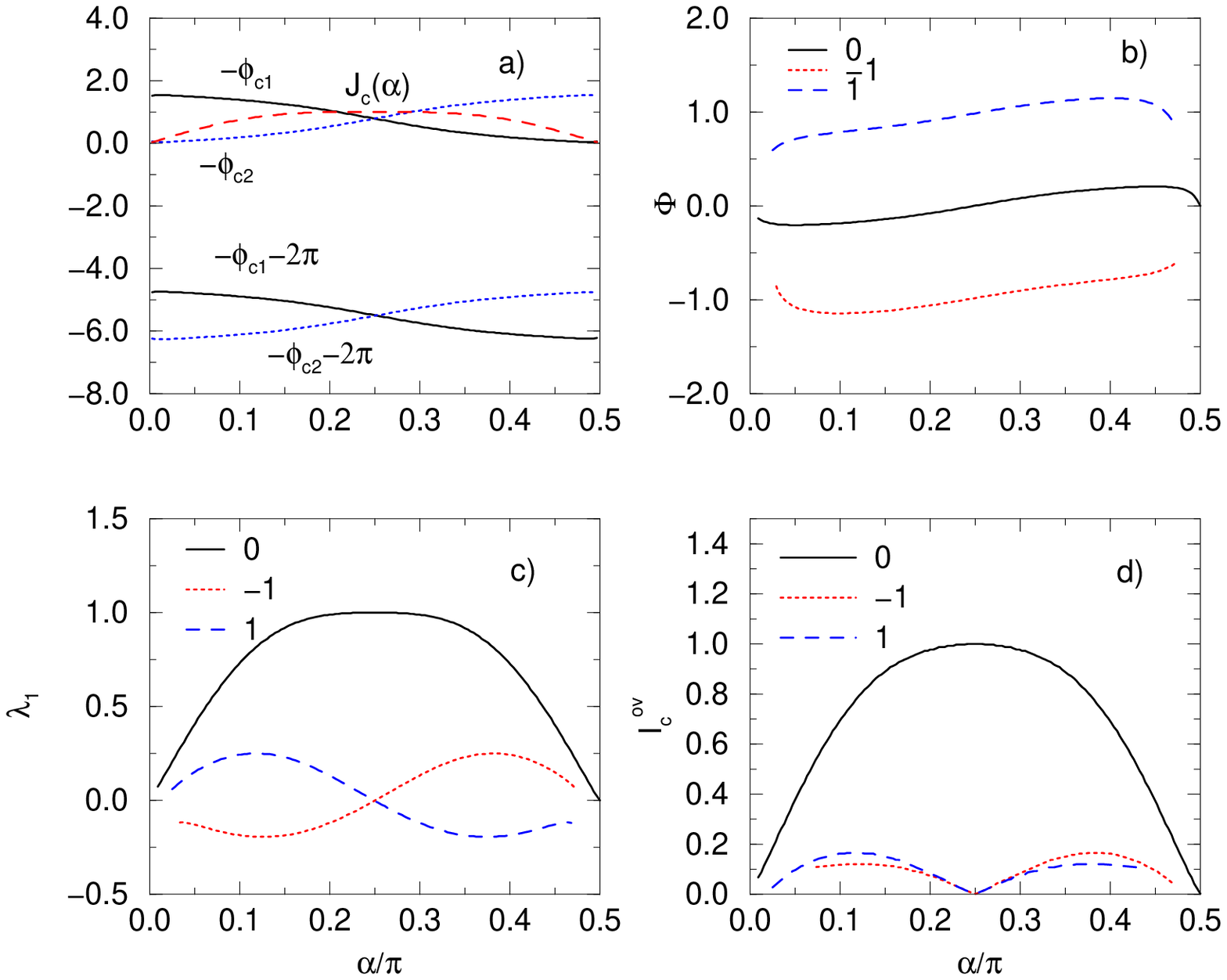,width=17cm,angle=0}}
\caption{
The same as in Fig. 2.
The pairing symmetry of the triplet 
superconductor is nodal $p$ wave.
}
\label{nodalp.fig}
\end{figure}

For the nodal $p$-wave case the $\phi_{c1}, \phi_{c2}$ 
are interchanged compared to the 
isotropic $p$-wave and $B_{1g}\times E_u$ cases 
and also their linear form is lost.
As a result the slope of the magnetic flux versus $\alpha$ seen in 
Fig. \ref{nodalp.fig}(b) has changed from negative in the 
isotropic $p$-wave and $B_{2g}\times E_u$ cases to positive here.
Now mode $1$ is stable in the interval $0<\alpha<\pi /4$ 
and mode $-1$ is stable in the interval $\pi /4<\alpha<\pi /2$, 
contrary to the isotropic $p$-wave case as seen in Fig. \ref{nodalp.fig}(c). 
The Josephson critical current density $\widetilde J_c$ has 
the same nodal form as in the $B_{2g}\times E_u$ case, with 
nodes at $0, \pi /2$, as seen in Fig. \ref{nodalp.fig}(a), since the nodal 
form of the order parameter is the same.   
The critical current for mode $0$
versus $\alpha$ follows the same form as in the 
$B_{2g}\times E_u$ case and it is zero at $0$, $\pi /2$ and 
maximum at $\alpha=\pi /4$, where the phase is uniform. 
Also the variation of the flux versus $\alpha$ 
seen in Fig. \ref{nodalp.fig}(b) is quite close to zero 
for mode $0$, 
and as a consequence 
the $I_c^{ov}$ vs $\alpha$ for mode $0$ is more broadened 
compared to the corresponding mode in the $B_{2g}\times E_u$ case. 
The mode $-1(1)$ 
becomes zero 
at $\alpha=\pi /4$ where a full antifluxon (fluxon) enters the 
junction. 
For $\alpha< \pi /4$, 
mode $1$ has less magnetic flux in absolute value
than does mode $-1$
and its critical current $I_c^{ov}$ is higher.
On the other hand for $\alpha>\pi /4$, 
mode $1$ has more magnetic flux in absolute value
than does mode $-1$
and its critical current $I_c^{ov}$ is lower.
Note that modes $-1,1$ stop abruptly at the angle where 
the variation of the flux is more abrupt  

\subsection{Nodeless $p$ wave}
\begin{figure}
 \centerline{\psfig{figure=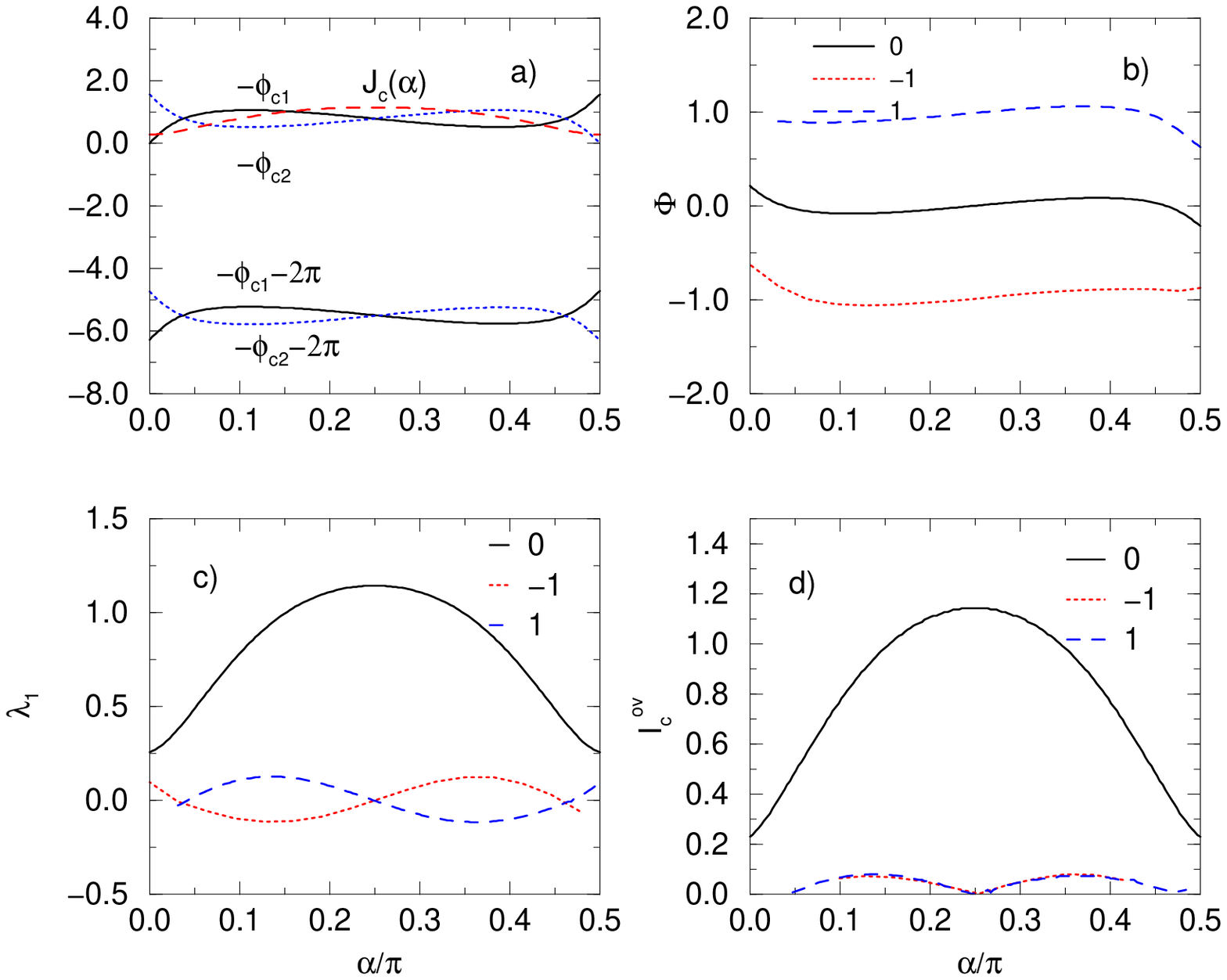,width=17cm,angle=0}}
\caption{
The same as in Fig. 2.
The pairing symmetry of the triplet 
superconductor is nodeless $p$ wave.
}
\label{MN.fig}
\end{figure}

For the nodeless $p$-wave case the variation of $\phi_{c1}, \phi_{c2}$ is 
more complex. It is similar to the nodal $p$-wave case except at values 
close to $0,\pi /2$ where it changes slope. 
As a result the magnetic flux versus $\alpha$ varies more 
closely to integer values and is almost flat 
as seen in Fig. \ref{MN.fig}(b). 
Also within the interval $0 < \alpha < \pi/2$ modes 
$-1,1$ change from stable to unstable for more than one 
time as seen in Fig. \ref{MN.fig}(c) while mode $0$ 
is stable for all values of $\alpha$. 
The $\widetilde J_c$ has a nodeless form as seen in Fig. \ref{MN.fig}(a),
i.e., it has no zero values. 
As a consequence the critical current for
mode $0$ is finite over the angles $\alpha$ and has its maximum values at 
$\alpha=\pi /4$ where the phase is uniform. 
At this point a full fluxon or antifluxon enters the junction 
and the critical current for modes $-1,1$ is zero. Also the magnetic flux 
for modes $-1,1$ varies close to values $-1,1$ and as a consequence their 
critical currents are almost coincident. 

\section{3d order parameters with horizontal lines of nodes}

In the preceding sections we considered purely 
two-dimensional order parameters. In this section we consider the 
case where the order parameter is three dimensional.
Recent measurement of thermal conductivity 
under a rotating in-plane magnetic field by Izawa {\it et al.}
\cite{izawa}
has shown a fourfold pattern with a minimum 
when $H \parallel [110]$ and
suggests that the nodes of the gap are on the
line with $k_z$=const as discussed in Ref. \cite{hasegawa}.
So the question arises also whether the results obtained in the
present paper remain valid for such parings. 

Several three-dimensional order parameters have 
been proposed in the literature:
Hasegawa {\it et al.} \cite{hasegawa} have proposed the 
pairing states with order parameter 
${\bf d}=\Delta_0\hat{\bf z}(k_x+ik_y)\cos(ck_z)$ and  
${\bf d}=\Delta_0\hat{\bf z}[\sin(ak_x/2)+i\sin(a/2)]\cos(ck_z/2)$
with $c$ being the lattice constant along the
$c$ axis. These states have horizontal lines of nodes 
at $k_z=\pm \pi/2c$
and $k_z=\pm \pi/c$, respectively,
and break the time-reversal symmetry.
Also Won and Maki \cite{won} have proposed an $f$-wave model 
with a superconducting order parameter ${\bf d}=\Delta_0\hat{\bf z}(k_x+ik_y)^2k_z$.
This state has horizontal lines of nodes where the basal plane $k_z=0$ 
intersects the Fermi surface.

In all the above pairing states the Josephson current within the
$ab$ plane is canceled, since the order parameter changes sign 
in the $k_z$ direction symmetrically around zero.
However it is possible to have finite Josephson current when
the Fermi surface is not exactly cylindrical but is corrugated as observed 
by the angle dependence of the de Haas-van Alphen effect \cite{yoshida}. In 
that case the cancellation would not be complete. 
Another possibility to have finite Josephson current 
within the $ab$ plane is to have a three-dimensional order parameter 
that is mixed with a two-dimensional one as suggested 
also in Ref. \cite{hasegawa}.
Then for the Josephson effect within the $ab$ plane, 
discussed in preceding sections, 
the quasiparticles will experience an 
effective order parameter that is quasi-two-dimensional and
the modulation of the critical current and magnetic flux with the 
misorientation angle within the 
$ab$ plane will show the nodal profile of the order parameter 
within the $ab$ plane discussed previously.

In an alternative geometry, 
the modulation of the critical current and spontaneous flux with the polar
misorientation angle within the $ac$ plane, i.e., when the azimuthal angle is
set to zero $\alpha=0$,  would reveal the
angular anisotropy of the order parameter along the $c$ axis.
In that case the effective order parameter felt by the quasiparticles 
is two dimensional. Moreover for that orientation the two-component 
order parameter is reduced to a one-component one which has 
a sign change but does not break time-reversal symmetry.

\begin{figure}
 \centerline{\psfig{figure=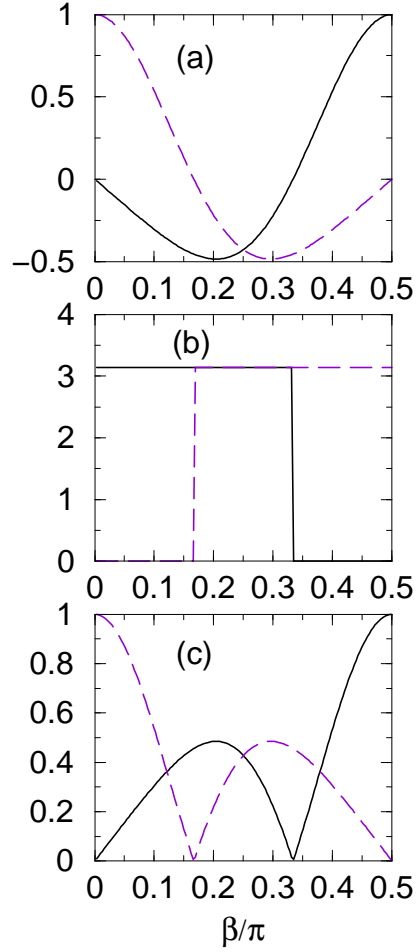,width=5.5cm,angle=0}}
\caption{
(a) The effective two-dimensional order parameter 
${\bf d}=\hat{\bf z}\sin \beta \cos(\pi \cos \beta )$
versus the polar angle $\beta$ for adjacent junction edges as 
solid and dashed lines. 
(b) The characteristic phase $\phi_{c1}$ ($\phi_{c2}$) as 
a solid (dashed line).
(c) The Josephson critical current densities $\widetilde{J}_{c1}$, 
$\widetilde{J}_{c2}$ for adjacent junction edges as 
solid and dashed lines.
}
\label{3d.fig}
\end{figure}

For example we consider the three-dimensional order parameter 
${\bf d}=\Delta_0\hat{\bf z}(k_x+ik_y)\cos(ck_z)$. For a spherical Fermi 
surface and $\alpha=0$, it is reduced to the two-dimensional order 
parameter ${\bf d}=\Delta_0\hat{\bf z}\sin \beta \cos(\pi \cos \beta )$.
We plot in Fig. \ref{3d.fig}(a) the order parameter 
for adjacent crystal edges versus the polar angle $\beta$.
It is seen that the order parameter changes sign, in a restricted angle regime,
at adjacent faces of the
crystal. Also the characteristic phases $\phi_{c1},\phi_{c2}$ differ by 
$\pi$ as seen in Fig. \ref{3d.fig}(b).
That would, for example, lead to
half-integer spontaneous flux states.
Note that the
Josephson critical current densities as seen in Fig. \ref{3d.fig}(c)
are not equal for adjacent junction edges.
However the asymmetry in the current densities will not affect
that spontaneous flux.
The modulation of the current with the polar angle
within the $ac$ plane will be highly anisotropic
for the three-dimensional order parameters. It will be suppressed
for $ck_z=\pi/2$ for the pairing state $\Delta \propto \cos ck_z$.

Since all states considered here have ${\bf d}\parallel\hat{\bf z}$
the Josephson current along the $c$ axis between the singlet 
and triplet superconductors is not allowed \cite{larkin}.
In fact the spin-orbit coupling strength of the triplet
superconductor would give rise to the orientational
dependent Josephson coupling between the SRO
and the conventional $s$-wave superconductor which completely
vanishes along the $c$ axis.
However the Josephson coupling is possible when the ${\bf d}$ vector is not 
exactly parallel to the current.
In that case we restrict to the corner junction in the $ac$ plane 
where the $c$ axis is tilted within the $ac$ plane for certain
polar angle.
For the pairing state $\Delta \propto k_z$ the 
singlet-superconductor/triplet-superconductor/singlet-superconductor 
junction along a direction slightly 
different from the $c$ axis can be used to probe the pairing symmetry 
from the sign change of the Josephson current at opposite faces 
of the junction \cite{geshkenbein}.
  
\section{magnetic field}
We now examine the influence of the magnetic field on the spontaneous
vortices,  for the $ab$-plane Josephson effect, 
for fixed junction orientation $\alpha=0$, 
where most experiments on corner junctions have been 
performed.
The critical current is finite, 
at $\alpha=0$ 
for the pairing states isotropic $p$ wave and 
$B_{1g} \times E_u$,  
while it is zero or very suppressed for the other pairing states. 
Hence we will restrict ourselves only to the isotropic $p$ wave 
and $B_{1g} \times E_u$ at $\alpha=0$ in this section.  
The modes extend to a large interval of the magnetic field, 
which is almost the same.
The instability at the extremum values of the magnetic field is due to the interaction of the flux entered from the edges with the intrinsic flux.  
The separation of the modes in magnetic flux is kept constantly equal 
to $\Phi_0$ as in the $s$-wave case, but is displaced compared 
to the $s$-wave case by an amount that is equal to the intrinsic flux.

\begin{figure}
\centerline{\psfig{figure=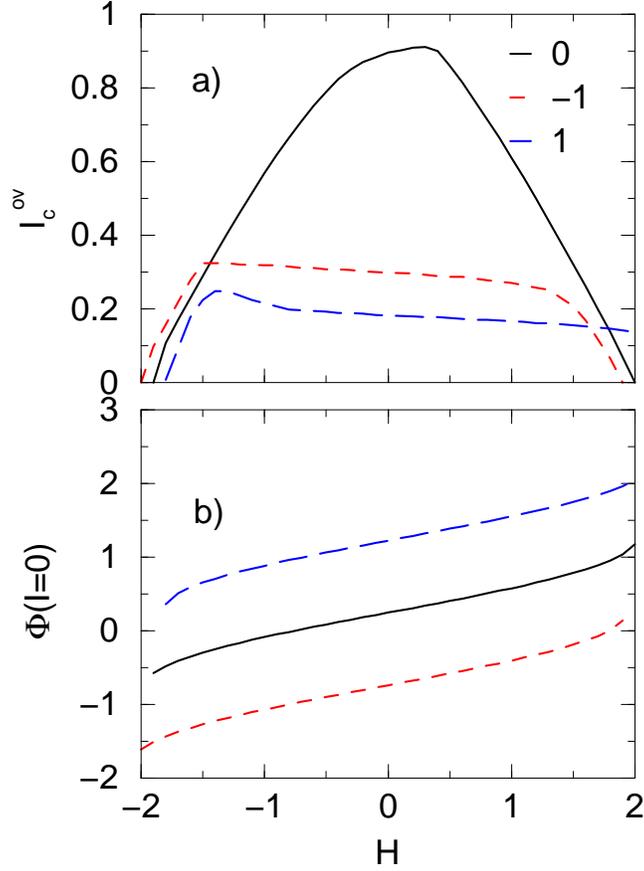,width=8.5cm,angle=0}}
\caption{
(a) The variation of the overlap $I_c^{ov}$ critical current with the magnetic field $H$ for the three vortex solutions $n=0,1,-1$. 
(b) The spontaneous magnetic flux $\Phi$ as a function of the 
magnetic field $H$, for the $0,-1,1$ mode solutions, 
for a junction of length 
$L=10$, and the angle $\alpha=0$. 
The pairing symmetry of the triplet 
superconductor is isotropic $p$ wave.
}
\label{isoH.fig}
\end{figure}

We plot in Fig. \ref{isoH.fig}(a) the overlap critical current $I_c^{ov}$ 
versus the magnetic field $H$, for a junction with length 
$L=10$, for angle $\alpha=0$, and for the isotropic $p$-wave case. 
The mode $0$, at $H=0$, contains positive flux $\Phi=0.25$ and has 
greater critical current than mode $-1$, which at $H=0$ contains 
negative magnetic flux and is equal to $-0.75$. 
So the mode with positive magnetic flux corresponds to the
highest critical current.
The maximum $I_c^{ov}$ for mode $0$ occurs to the right of $H=0$.
Note that the displacement of modes $0,-1,1$ 
in $\Phi$ from their positions in the $s$-wave case is small
and thus the $I_c^{ov}$ vs $H$ diagram is similar to the 
$s$-wave case.   

\begin{figure}
\centerline{\psfig{figure=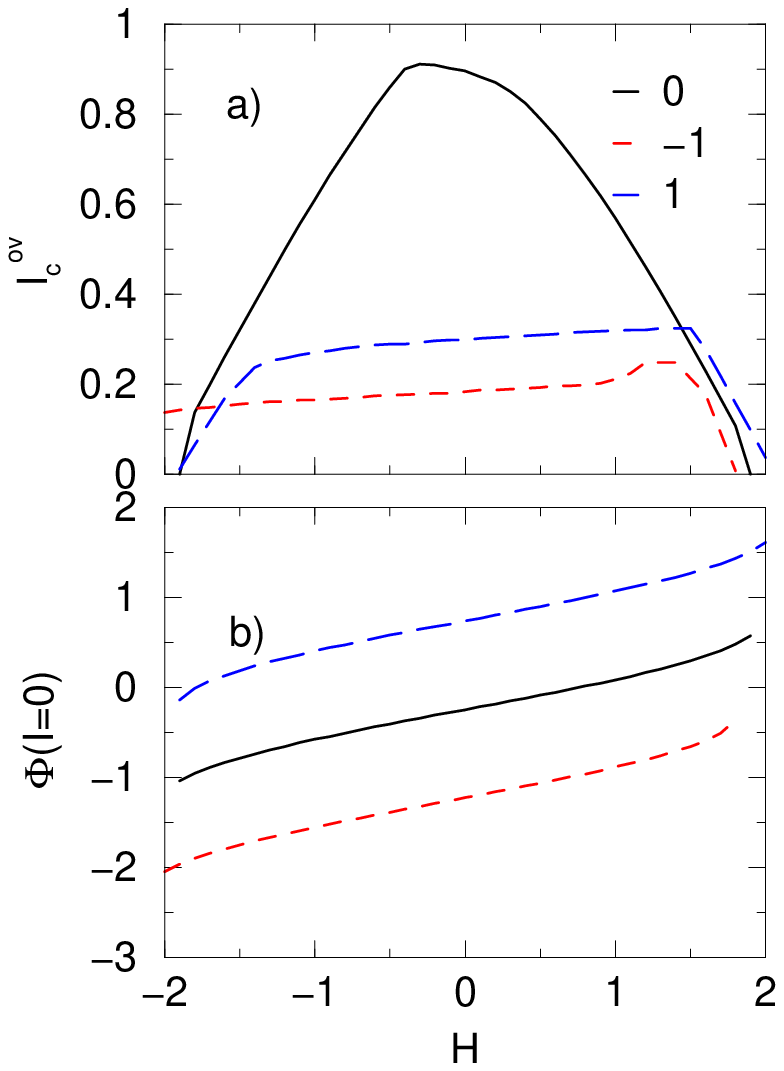,width=8.5cm,angle=0}}
\caption{
The same as in Fig. 10. The pairing symmetry of the triplet 
superconductor is $B_{1g}\times E_u$.
}
\label{cosH.fig}
\end{figure}

In Figs. \ref{cosH.fig}(a) and \ref{cosH.fig}(b), 
the overlap critical current $I_c^{ov}$
versus the magnetic field $H$ and the spontaneous flux vs 
magnetic field is plotted for the $B_{1g}\times E_u$ case.
The mode $0$, at $H=0$, contains negative flux $\Phi=-0.25$ and has
greater critical current than mode $1$, which at $H=0$ contains
positive magnetic flux and is equal to $0.75$. So differently from the 
isotropic $p$-wave case, 
the mode with negative magnetic flux corresponds to the 
highest critical current.

For the rest of the pairing states the 
junction orientation $\alpha=0$ gives zero or
relatively small values 
for the Josephson critical current density $J_c(\alpha)$,
which means that the critical current the junction can carry
will be zero or very suppressed 
compared to the previous cases.
For the $ac$-plane Josephson effect for three-dimensional order
parameters it is possible to have a
sign change of the order parameter at adjacent faces of
the corner junction that would, for example lead to an
anomalous Fraunhofer pattern with "dip" at zero magnetic field.
The asymmetry in the Josephson critical current densities 
will only cause the "dip" to be shallow and will maintain the 
symmetry of the pattern.
\section{Experimental relevance} 

For the $ab$-plane Josephson effect for two-dimensional order 
parameters, the measurement of the magnetic interference pattern at $\alpha=0$ 
where the $a$ and $b$ crystallographic axes are perpendicular to the 
junction's faces will not give any finite value for the critical current 
$I_c^{ov}$ for the pairing states $B_{2g}\times E_u$ or the nodal $p$ wave 
since the value $\alpha=0$ is a node of the order parameter 
for these symmetries. 
For the nodeless $p$-wave case the Josephson critical 
current density is small at 
$\alpha=0$ and so is the critical current. So the observation of 
finite critical current $I_c^{ov}$ in corner junctions at $\alpha=0$ 
would be an indication for isotropic $p$-wave or $B_{1g}\times E_u$ 
pairing states where the value $\alpha=0$ is not a node.
Furthermore the suppression of
the critical currents of the nodal state
$B_{1g}\times E_u$  near the nodes, i.e., at $\alpha=\pm\pi/4$,
can be used to distinguish it from the
nodeless isotropic $p$-wave state, for which the critical current 
for no-flux mode $0$ does not become zero.

On the other hand for the $ac$-plane tunneling, the orientation where 
$ck_z=\pi/2$ will suppress the Josephson critical current 
for the pairing state $\Delta \propto \cos ck_z$ since it is 
a node for the order parameter. 
Also the singlet/triplet/singlet junction along the direction 
slightly different from the $c$ axis will probe the pairing state
$\Delta \propto k_z$ from the sign change of the Josephson current 
at opposite sides of the junction.

Another experiment that we propose here to discriminate between the candidate's 
pairing states consists of a SQUID between a Sr$_2$RuO$_4$ thin film patterned 
into a circle with a series of Nb-Au-Sr$_2$RuO$_4$ edge junctions spaced 
at a fixed angle step. 
The tunneling directions can be defined lithographically and patterned 
by ion milling of a $c$-axis-oriented film to study the $ab$-plane 
Josephson effect or a $b$-axis-oriented film to study the 
$ac$-plane Josephson effect. 
The measurement of the critical current versus the angle $\alpha$, 
which mainly probes the magnitude of the order parameter, is expected 
to show large anisotropy with the angle $\alpha$. 
An analogous experiment has already been performed in the SQUID geometry for 
YBa$_2$Cu$_3$O$_7$ thin films \cite{yanoff}.

A number of complicating factors are involved in the interpretation 
of the experiments with corner junctions that could modify the anticipated 
results. These include the asymmetry in the Josephson critical current 
densities, the flux trapping, and the sample geometry. 
Also the study of the direction-dependent Josephson effect will be 
influenced by the roughness of the interface due to the existence 
of Ru lamellas, and also by the difficulty in cleaning  and polishing 
a crystal at angles different than $0$ and $\pi/2$.
However these effects are not sufficiently large enough to change the 
qualitative picture of the experiments.     

\section{Conclusions} 
 
We followed the modulation of the vortices that are spontaneously 
formed in a corner junction between a triplet superconductor 
breaking time-reversal symmetry and an $s$-wave superconductor with 
the junction orientation angle.
The nodal structure of the order parameter is 
reflected in the Josephson critical current density, and influences 
the critical current versus $\alpha$ diagram and the magnetic flux of the
spontaneous vortex states that are formed at the corner junction.
The critical current vanishes at the angles where the order parameter
has nodes. 
In addition it becomes zero at the angles where a full fluxon 
or antifluxon enters the junction.
For the nodeless pairing state mode $0$ does not 
vanish, and has the nodal profile of the corresponding 
order parameter. 
The presence of spontaneous magnetic flux can be used to distinguish 
between different symmetry states.

The external magnetic field for 
the junction orientation $\alpha=0$ where the $a$ axis points 
towards the junction edge will only influence
the states in which their order parameter is finite at 
$\alpha=0$, i.e., isotropic $p$ wave and $B_{1g}\times E_u$.  
The critical current and the effect of the magnetic 
field will be zero or very suppressed 
for the other cases.
For three-dimensional order parameters the Josephson critical 
current modulation, which mainly probes the magnitude of the 
order parameter
with the polar angle, is expected to be highly anisotropic 
for the $ac$-plane 
Josephson effect.

In corner junctions with broken time-reversal symmetry, the spontaneus 
flux and the magnetic 
interference pattern, even in the short junction limit, are influenced 
by the magnitude of the order parameter and also 
the difference in the intrinsic phases $\phi_{c1}$ and $\phi_{c2}$, 
at adjacent junction edges. 
This difference  
uniquely 
characterizes the pairing symmetry and can be used without 
any ambiguity to probe anisotropic pairing states with 
sign change of the order parameter between orthogonal directions 
in $k$ space.

\newpage

\begin{table}
\caption{
The magnetic flux ($\Phi$) in terms of $\phi_{c1}$, $\phi_{c2}$
for the spontaneous solutions that exist in a corner junction
between a triplet superconductor with time-reversal broken symmetry
and an $s$-wave superconductor ($\phi_{c1}, \phi_{c2}$
is the extra phase difference in the two edges of the corner junction due to the
different orientations, of the $a$ axis of the triplet superconductor). 
We present only the minimum
flux states $n=0,-1,1$. 
}
\begin{tabular}{rc}
Vortex state & Magnetic flux ($\Phi$)\\ \tableline
$0$ & $(-\phi_{c2}+\phi_{c1})/2\pi$\\
$1$ & $(-\phi_{c2}+\phi_{c1}+2\pi)/2\pi$\\
$-1$ & $(-\phi_{c2}+\phi_{c1}-2\pi)/2\pi$\\
\end{tabular}
~~~\label{n=01_1}
\end{table}

\begin{table}
\caption{
The magnetic flux ($\Phi$) in terms of $\alpha$
for the spontaneous vortex states that exist in a corner junction
between a triplet superconductor with isotropic $p$-wave symmetry
and an $s$-wave superconductor. 
}
\begin{tabular}{rc}
Vortex state & Magnetic flux ($\Phi$)\\ \tableline
$0$ & $0.25-\alpha/\pi$\\
$1$ & $1.25-\alpha/\pi$\\
$-1$ & $-0.75-\alpha/\pi$\\
\end{tabular}
~~~\label{isotable}
\end{table}

\end{document}